\documentclass[jgr]{agu2001}
\usepackage{graphicx}
\lefthead{OUILLON, DUCORBIER and SORNETTE}
\righthead{3D determination of fault patterns}	
\authoraddr{Guy Ouillon, Lithophyse, 1 rue de la croix, 06300 Nice, France 
(e-mail: lithophyse@free.fr)}
\authoraddr{Caroline Ducorbier, LPMC, CNRS and University of Nice, 06108 Nice, France ()}
\authoraddr{Didier Sornette, D-MTEC, ETH Zurich, Kreuzplatz 5,
CH-8032 Zurich, Switzerland and D-ESS and IGPP, UCLA, Los Angeles, California, USA
(e-mail: dsornette@ethz.ch)}

\begin{document}
\title{Automatic Reconstruction of Fault Networks from Seismicity Catalogs: 3D Optimal Anisotropic Dynamic Clustering}
\author{G. Ouillon$^1$, C. Ducorbier$^2$ and D. Sornette$^{3,4}$}
\affil{$^1$ Lithophyse, 1 rue de la croix, 06300 Nice, France \\
$^2$  Laboratoire de Physique de la Mati\`{e}re Condens\'{e}e, CNRS UMR 6622,
Universit\'{e} de Nice-Sophia Antipolis, Parc Valrose, 06108 Nice, France\\
$^3$ D-MTEC, ETH Zurich, Kreuzplatz 5, CH-8032 Zurich, Switzerland\\
$^4$ Department of Earth and Space Sciences and
Institute of Geophysics and Planetary Physics, 
University of California, Los Angeles, California 90095-1567}

\newcommand{\be}{\begin{equation}}
\newcommand{\ee}{\end{equation}}
\newcommand{\ba}{\begin{eqnarray}}
\newcommand{\ea}{\end{eqnarray}}
\newenvironment{technical}{\begin{quotation}\small}{\end{quotation}}

\begin{abstract}
We propose a new pattern recognition method that is able to reconstruct
the $3D$ structure of the active part of a fault network using the
spatial location of earthquakes. The method is a generalization of the
so-called {\it dynamic clustering} method, that originally partitions a
set of datapoints into clusters, using a global minimization criterion
over the spatial inertia of those clusters. The new method improves on
it by taking into account the full spatial inertia tensor of each
cluster, in order to partition the dataset into fault-like, anisotropic
clusters. Given a catalog of seismic events, the output is the optimal
set of plane segments that fits the spatial structure of the data. Each
plane segment is fully characterized by its location, size and
orientation. The main tunable parameter is the accuracy of the
earthquake localizations, which fixes the resolution, {\it i.e.} the
residual variance of the fit. The resolution determines the number of
fault segments needed to describe the earthquake catalog, the better the
resolution, the finer the structure of the reconstructed fault segments.
The algorithm reconstructs successfully the fault segments of synthetic
earthquake catalogs. Applied to the real catalog constituted of a subset
of the aftershocks sequence of the 28th June 1992 Landers earthquake in
Southern California, the reconstructed plane segments fully agree with
faults already known on geological maps, or with blind faults that
appear quite obvious on longer-term catalogs. Future improvements of the
method are discussed, as well as its potential use in the multi-scale
study of the inner structure of fault zones. 
\end{abstract}

\begin{article}

\section{Introduction}

Tectonic deformation encompasses a wide spectrum of scales, both in
spatial and temporal dimensions. This certainly constitutes a
fundamental obstacle to thoroughly study, from observations alone, the
whole set of relevant processes. 
It is now clear that seismic events occur on faults, while faults grow
by accumulation of slip, by the growth of damage in the region of their
tips, and by linkage between pre-existing faults of various sizes, all
three processes occuring in large part during earthquakes (see {\it
Scholz (2002)} for a review of earthquake nucleation and fault growth
mechanisms and {\it Sornette et al. (1990)} and {\it Sornette (1991)} for a general
theoretical set-up). In the last twenty years, a large amount of
knowledge has been obtained on the statistical properties and general
phenomenology of faults and earthquakes (including the distribution of
sizes, of inter-event time distributions, the geometrical correlations,
and so on), but a clear physical and mechanical understanding of the
links and interactions between and among them is still missing. Such
knowledge may prove decisive to drastically improve our understanding of
seismic risks, especially the time-dependent risks associated with the
future large and potentially destructive seismic events.

The study of natural fault networks and of earthquake catalogs is
generally conducted in two very different ways, which clearly reveal
still distinct scientific cultures. On one side, brittle tectonics
(within which we include seismotectonics) and structural geology pays a
lot of attention to structures, but dwells quite little on their nonlinear
collective dynamics. Thus, brittle tectonics and structural geology are
incapable of any predictive seismological statement, despite an
impressive amount of theoretical and laboratory-validated concepts
coming from rock physics and rheology (see {\it Scholz (2002)}, {\it
Passchier and Trouw (2005)} and {\it Pollard and Fletcher (2005)}), as
well as from observational techniques. Brittle tectonics and structural
geology mainly consider structures as isolated and thus focus on the
``one-body'' problem, generally solvable through the use of classical
physics and/or the use of the mechanics of homogeneous media. On the
other side, statistical analyses can reveal patterns, but rarely
provide anything else than purely empirical descriptions of the data.
Some statistical analyses may even be seriously flawed due to their very
nature (see for instance {\it Ouillon and Sornette (1996)} who reveal
a mechanism leading to spurious multifractal analyses of fault patterns). For instance, when
dealing with earthquakes, most spatial analyses deal with epicenters or
hypocenters. Since real earthquakes involve extended oriented
structures, statistical analyses based on point-wise representation of
earthquakes may led to spurious correlations between events, especially
the smallest ones, which may distord any model of the collective
build-up of a large shock by previous smaller events.

Statistical analyses can also be performed using fault networks, but
they are most often restricted to the available 2D surface maps. It is
then difficult to assess the amount of information that is genuinely
representative of 3D active faulting at depth and, as such, related to
seismogenesis, in the goal of developing better understanding and
improved forecasts. There are several problems when working with
networks of surface-intersecting faults. First, the sampling of fault
data is by far more difficult and time-consuming than constructing
earthquake catalogs, with the often unavoidable existence of subjective
inputs. Second, one can only map small-scale features in small-scale
domains (see {\it Ouillon et al (1995)} and {\it Ouillon et al (1996)}
for a rather unique high-resolution multi-scale analysis of fracturing
and faulting of the Arabic plate). The domains chosen for such sampling
may not be good representative, as their choice may either be arbitrary
or reflect practical constraints. Third, there can exist some masks
(like cities, deserts or vegetation) which make observations impossible
in sometimes vast areas. This can seriously bias statistical results
(see {\it Ouillon and Sornette (1996)}). In any case, if one takes
account of all those shortcomings, spatial analyses are possible at any
scale (from millimeters to thousands of kilometers - see {\it Ouillon et
al (1995)} and {\it Ouillon et al (1996)} which showed how to exploit
such data with multifractal and anisotropic wavelet methods so as to
reconstruct objective fault maps at multiple scales), but there is a
terrible lack of data in the time domain. 
{\it Darrozes et al.  (1998)} and {\it Gaillot et al.  (2002)} studied the 
aftershock sequence of the 1980 M5.1 Arudy earthquake in the 
French Pyr\'en\'ees using anisotropic wavelets. Their method is 
an extension of the Optimal Anisotropic Wavelet Coefficient method 
({\it Ouillon et al (1995);  Ouillon et al (1996)}). This method uses 
anisotropic filters, which properties (size and orientation) can 
vary from one location to another within the same dataset.
 At each spatial location, the shape of the filter is chosen such 
 that it reveals best the local anisotropic features, if any 
 (hence the ``optimal'' qualification). In this way, each pixel in 
 the image is characterized by a local optimal anisotropy. 
 This method allowed one to detect and map linear structures 
 in the Arudy sequence, and to quantify their orientation. 
 However, identified faults are still sets of neighbouring pixels, 
 each pixel ``ignoring'' that it possibly belongs to the same 
 fault as its neighbours. The images provided by wavelet 
 analysis still need to be post-processed in order to compute, 
for example, the size of the lineaments. The method 
is also restricted to 2D datasets. Extension to 3D geometries 
would be theoretically straightforward, but the size of images 
to handle at the scale of a regional catalogue is totally unrealistic, 
as large volumes of space would have to be discretized at scales 
much lower than the earthquake localization accuracy.

The structure of the present
state of a fault pattern integrates its whole history, {\it i.e.}
reflects the largest possible time scale, which can be several tens of
millions of years. At very short time scales, seismology offers pictures
of the slip pattern over a reconstructed fault surface that occured
during an earthquake. But this kind of inversion is performed only for
the largest shocks. In between, paleo-seismology allows one to decipher
the sliding history along discontinuities, but it still focuses on major
faults and major seismic events for obvious reasons of time and space
resolution. In fact, there is absolutely no work at all on the mechanics
of faulting within a fault network as a whole derived from field
observations, at time scales comparable with those of a seismic catalog.
Obtaining such information would certainly constrain significantly our
concepts and methods on the space-time organization of earthquakes and
on the potential for their forecasts, with the ultimate goal of better
assessment of time-dependent hazard.

Most studies focus on earthquake catalogs. This comes from the fact that
earthquake catalogs contain in general a large number of events,
allowing accurate statistics to be computed, and that building such a
database is reasonably mastered at the technical level. A homogeneous
seismographic network can span a very large domain, which allow for
spatial analyses at scales from a few tens of meters to hundreds or
thousands of kilometers, at least in $2D$. Depth is often omitted in
most statistical analyses, as the increases of pressure and temperature,
as well as the finite thickness of the seismogenic layer, clearly break
any possible symmetry along the vertical direction, which one generally
does not know how to deal with when performing statistics (see however
{\it B\'ethoux et al (1998)} for a genuine 3D wavelet analysis of
seismicity in the Alpine arc). Another reason for the use of seismic
catalogs for the analysis of fault networks is that they also allow for
clean analyses in the time domain, at scales varying from few seconds to
decades.

Up to now, most of the information available on the structure of fault
patterns comes from surface mapping at various scales. One can however
question the reliability of such data sets, as earthquakes catalogs
generally include large quantities of events which do not seem to be linked
to any such faults, but rather seem to be associated with active faults buried at depth which do
not cross-cut the surface. It is thus rather clear that comparing $2D$
map views of the long-term cumulative brittle deformation to the $3D$
structure of the short-term incremental deformation can provide at best a
poor insight into, and at worst a  biased incorrect model of, 
the multi-scale tectonic processes as a whole. A detailed
knowledge of the $3D$ structure of fault networks is needed to
better understand the mechanics of earthquake interaction and collective
behaviour. In the last few years, in parallel to the ever increasing
capabilities of computers, earthquake localization algorithms have
significantly improved so that the accuracy for the spatial
resolution of earthquake hypocenters dropped down to about $1km$, and
even to a few tens of meters within clusters (when using relative
locations through cross-correlating waveforms, {\it Shearer et al
(2005)}). This provides new opportunities to learn about the detailed
$3D$ structure of fault patterns at depth using seismicity itself as a
diagnostic.

The basis of our paper is that seismic hazard assessment faces 
two fundamental bottlenecks: (i) catalogs are inherently incomplete with 
respect to the typical recurrence time of large earthquakes which are 
many times larger than the duration of the available catalogs; 
(ii) earthquakes occur on faults, but most of them are still unknown, 
drastically limiting our understanding and our assessment of seismic risks. 
The principal issue therefore lies in the association between earthquakes and faults. 
A recent effort to assign earthquakes and simple fault structures in the 
San Francisco Bay area showed significant disparities that arose from 
the simplified geometry of fault zones at depth and the amount and 
direction of systematic biases in the calculation of earthquake 
hypocenters ({\it Wesson, 2003}). The geometry of an active 
fault zone is often constrained by mapping the surface trace; dip 
angle at depth and depth extent are either constrained by results 
of controlled source seismology (if available) and the distribution of 
hypocenter locations or they are just extrapolated using geometric 
constraints, if seismologic constraints are not available. For example, 
one of the most sophisticated fault models available, the Community 
Fault Model (CFM) of the Southern California Earthquake Center (SCEC), 
combines available information on surface traces, seismicity, seismic 
reflection profiles, borehole data, and other subsurface imaging techniques 
to provide three-dimensional representations of major strike-slip, blind-thrust, 
and oblique-reverse faults of southern California ({\it Plesch, 2002}). 
Each fault is represented by a triangulated surface in a precise 
geographic reference frame. However, the representation of a fault 
by a simple surface cannot reflect the fine-detailed structure seen in
extinct fault zones and in drilling experiments of active faults
(e.g. {\it Scholz, 2002} ; {\it Faulkner, 2003}). These results 
suggest that fault zones actually consist of narrow earthquake-generating 
cores, possibly accompanied by small subsidiary faults. Our present 
contribution is to propose a general method to identify and locate 
active faults in seismically active regions by using a scientifically 
rigorous approach based on active seismicity. Our goal is to gain 
a better understanding of the link between fault structures and earthquakes. 

For this, the present paper presents a new pattern recognition method which determines an
optimal set of plane segments, hereafter labeled as `faults' that best
fit a given set of earthquake hypocenters. In the first part, we provide
a short overview of pattern recognition methods used to detect linear or
planar features in images, such as the Hough and wavelet transforms. We
then present in more details the dynamic clustering method, which
provides a partition of a set of data points into clusters, before
detailing our new method, a generalization of dynamic clustering to the
case of anisotropic structures. We illustrate this new technique through
benchmark tests on synthetic datasets, before providing results obtained
on a subset of the Landers aftershocks sequence. The final discussion of
our results compares this new method with other clustering methods, and
outline its potential use in the study of fault-zones structures. We
then sketch the next set of improvements for this method which will be
developed in sequel papers.

\section{Line and plane detection in image analysis}

The detection of linear and planar structures in seismotectonics has a
long history, but still suffers from a lack of quantitative methods.
From the early years of instrumental seismology, the main instrument to
identify faults from earthquakes has indeed been the eye. It is by
visual inspection that tectonic plates boundaries or the Benioff plane
in subduction zones have been delineated. Visual inspection remains the
main approach to identify blind faults at smaller scales. Very few
efforts have been devoted to the development of automatic digital
detections of linear or planar spatial features from earthquake catalogs.
We shall thus now describe two major approaches in image analysis, and
outline their advantages and drawbacks, before turning to dynamic clustering.

\subsection{The Hough transform}

The Hough transform (see {\it Duda and Hart, 1972}) is a technique that
is widely used in digital image analysis, with applications, for
example, in the detection and characterization of edges, or the
reconstruction of trajectories in particle physics experiments. The
original Hough transform identifies linear features within an image, but
it can be extended to other shapes (like circles or ellipses). We shall
anyway present it for straight line detection in $2D$ for the sake of
simplicity. The basic idea is that, given a set of data points, there is
an infinite number of lines that can pass through each point. The aim of
the Hough transform is to determine the set of lines that get through
several points.

Each line can be parameterized using two parameters $r$ and $\theta$ in
the Hough space. The parameter $r$ is the distance between this line and
the origin, and $\theta$ is the orientation of the normal to that line
(and is measured from the abscissa axis). Then, the infinite number of
lines passing through a given point defines a curve $r(\theta)$ in the
Hough plane. All lines passing through a point located at $(x_0,y_0)$
obey the equation $r(\theta) = x_0 \cdot \cos(\theta) + y_0 \cdot
\sin(\theta)$. If we now consider several points of our dataset that are
located on the same straight line in direct space, their associated
$r(\theta)$ curves will all cross at the same $(r,\theta)$ point in the
Hough space. Reciprocally, the corresponding line is then fully identified.
Note that several distinct sets of $r(\theta)$ curves can cross in the
Hough space at different $(r,\theta)$ locations, so that as many linear
features can be extracted at once. This idea can be extended to
arbitrary $2D$ structures (such as circles or ellipses) or to 3D spaces
and planar features, provided that one extends the dimension of the
Hough space correspondingly. For example, this dimension is equal to $4$
when one wants to detect straight lines in a $3D$ real space, so that
the method is not used in practice in that case (or rather in a much
simplified form, provided there is only a very small number of such
lines to detect, see {\it Bhattacharya et al (2000)}). The dimension of
the Hough space drops to $3$ when dealing with planar features.

The main problem of this technique is that it does not provide directly
the spatial extension of the linear or planar features, only their
positions in space, as each line (or plane) has an infinite size. A
post-processing step must then be performed in order to compute the
finite extension. Another drawback is that the Hough transform does not
take account of the uncertainties in the localization of the set of data
points, a crucial parameter in seismology. All in all, it appears that
the Hough transform is very efficient only when dealing with few very
clean ordered patterns (see however some specific strategies that can be
defined to deal with real fracture planes ({\it Sarti and Tubaro,
2002}). Another major argument against the use of the Hough transform
for fault segment reconstruction from seismic catalogs is that there is
absolutely no way to include any other information one may have about
seismic events, such as their seismic moment tensors, focal mechanisms
or magnitudes.

\subsection{The Optimal Anisotropic Wavelet Coefficient method}

The characterization of anisotropy is a basic task in structural
geology, tectonics and seismology. The orientations of structures stem
from mechanical boundary conditions, which we are in general interested
to invert. They control the future evolution of the system, which we are
interested to predict. Tools for quantifying anisotropy generally
materialize in rose diagrams (for $2D$ fracture maps for instance) or in
stereographical projections (when data are sampled in $3D$ space). The
main drawback of these representations is that they are scale-dependent.
For example, as shown in ({\it Ouillon, 1995 ; Ouillon et al, 1995}), in
the case of {\it en-\'echelon} fractures, small-scale and large-scale
features will possess fundamentally different anisotropy properties. In
that case, mapping at small scales will not give any clue on the large
scale properties of the system. The so-called Optimal Anisotropic Wavelet
Coefficient (OAWC) method, presented in the above publications, was designed to
specifically address this problem.

For $2D$ signals, such as fault maps and fracture maps, a wavelet is a
band-pass filter which is characterized, in real space, by a width
(called resolution), a length and an azimut. In a nutshell, the OAWC
method consists in fixing the resolution and convolving the wavelet with
the map of fault traces, varying its length and azimut. For any given
position, when the result of the convolution (which is the wavelet
coefficient) reaches a maximum, one stores the associated azimut of the
wavelet and switch to the next spatial location. Building a histogram of
such optimal azimuts provides a rose diagram at the considered
resolution. Performing this process at different resolutions allows one
to describe the evolution of anisotropy with scale from a single initial
data set. Looking at various maps of fractures and faults patterns in
Saoudi Arabia, {\it Ouillon, 1995} and  {\it Ouillon et al, 1995} were able to
show that anisotropy properties change drastically at resolutions that
can be related to the thicknesses of the mechanical layers involved in
the brittle deformation process (from sandstone beds up to the
continental crust).

The OAWC method has been extended to characterize the anisotropy of
mineralization in thin plates, as well as to detect faults or lineaments
from earthquakes epicenter maps ({\it Darrozes et al, 1997 ; Darrozes et
al, 1998 ; Gaillot et al, 1997 ; Gaillot et al, 1999 ; Gr\'egoire et al,
1998}). In this second class of applications, the issue concerning the
finite accuracy of event localizations is addressed by considering a
wavelet resolution larger than the spatial uncertainties. However, this
method does not provide direct information about the size of the
structures, and indeed does not manipulate structures as such. Each
point in space is characterized by anisotropy properties, but the method
do not naturally construct clusters from neighbouring points with
similar characteristics. Moreover, its extension to $3D$ signals and
their associated 3D patterns would prove unreasonably time-consuming (as
space has to be discretized at a scale smaller than the chosen
resolution) and would suffer from edge effects near the top and bottom
of the seismogenic zone. Nothwithstanding its power and flexibility,
it is no better that the Hough transform for our present problem
of reconstructing a network of fault segments best associated with 
a given seismic catalog.

\section{Dynamic clustering}

\subsection{Dynamic clustering in 2D}

Dynamic clustering (also known as {\it k-means}) is a very general image processing technique 
that allows one to partition a set of data points using a simple criterion 
of inertia. The method is described in many papers or textbooks (see for example
{\it Mac Queen (1967)}), 
so it will suffice for our purposes to perform a practical demonstration in $2D$ -
 but extension of the method  to $3D$ is straightforward.

We first define a set of $N$ datapoints $(x_i,y_i)$, with $i=1,N$.
Fig.~\ref{nudyn_plate_1} (stage a) is an example with $N=9$.
Our goal is to partition this dataset into a given number $N_c$ 
of clusters, with $N_c<N$. In the example illustrated in 
Fig.~\ref{nudyn_plate_1},  $N_c=2$. The first step consists 
in throwing $2$ more points at random represented with square symbols (stage b)).
Note that those added points can be chosen to coincide with 
points belonging to the initial dataset.
We now link each circle to the closest square, and obtain a 
first partition of the dataset into $2$ clusters
(stage c)). The first cluster features $2$ datapoints, while the 
second cluster features $7$ data points in our example.
Each square is now moved to the barycenter of 
the associated cluster (stage d)), leaving the partition intact. 
As the squares change
position, so do the distances between squares and circles. We link
once again each circle to the closest square, 
which updates the partition (stage e)). Clusters now feature 
respectively $4$ and $5$ data points.
Squares are then moved once again to the barycenters of the 
associated clusters (stage f)). 
We associate once more each circle to the closest square 
(stage g)), which updates the partition to a new one, the squares
being moved to the barycenters of the corresponding clusters. 
We obtain stage h (the signification of the dashed-line circles
will be given below). One can then check 
that any further iteration doesn't change the partition anymore
(that is, the location of the squares), so that the method converges
to a fixed structure, which is the final partition of the dataset.

For any given partition, we can compute the inertia of each 
cluster $i_c$ (featuring $N_{i_c}$ points) by:
\be
I_{i_c} = \frac{1}{N_{i_c}} \sum_{j=1}^{N_{i_c}} D_j^2
\ee
where $D_j$ is the distance between one point belonging to the cluster 
$i_c$ and the barycenter of that cluster.  It can then be proven
that the final partition we obtain minimizes the sum of inertia 
over all clusters. However, there can exist several local
minima for the global inertia in the partition space, so that one has 
to run the dynamic clustering procedure several times, 
using different initial location of the squares in order to select the 
partition which corresponds to the genuine global minimum.
Stage h shows the final partition we obtained starting from 
stage b, with $2$ dashed-line circles. 
Each circle is centered on the barycenter
of a cluster, and its radius is taken as twice the square 
root of the inertia of that cluster. The role of those circles
is just to illustrate the size of each cluster, and the way we quantify 
this size will become clearer later. The total variance
in stage h is $3.5375$ (note that coordinates are dimensionless in our example).

We performed this dynamic clustering process using 
hundreds of different initial positions for the squares, and were able to 
find two other local variance minima (with corresponding 
partitions illustrated on Fig. \ref{nudyn_plate_2}a and 
\ref{nudyn_plate_2}b). The final inertia of the configuration shown 
in Fig. \ref{nudyn_plate_2}a is $3.1111$, 
while the one we computed for Fig. \ref{nudyn_plate_2}b is $3.5833$. 
The partition which minimizes the global inertia is thus
the one shown on Fig. \ref{nudyn_plate_2}a. Note however 
that those three minima are not very different from each other,
so that all of them appear as very reasonable partitions. 
The similarity of the numerical values of the variances suggests 
that all three partitions would be barely discriminated by a 
human operator alone, which one generally refers to as `subjectivity'. 
Note also that the final partitions display data points that lay almost
half-way between two squares, so that adding some noise on 
the positions of the data points (to simulate earthquake localization uncertainties)
would certainly make a given final partition switch to one 
of the two others, and vice-versa.

\subsection{The problem of anisotropy in 3D - Principal component analysis}

Dynamic clustering, as described in the previous section, is an iterative 
procedure which minimizes the sum of variances 
over a given number of clusters. As presently set up, dynamical clustering
suffers from the two following drawbacks:
\begin{itemize}
\item[(i)] the user has to choose a priori the number of clusters necessary 
to partition our data set; the method to remove this constraint
will be addressed in  the next section;
\item[(ii)] Using a criterion in terms of the minimization of the total inertia
is certainly not adequate when data points have to be partitioned into anisotropic
 clusters, as we plan to do by associating earthquakes with faults. 
As an alternative, we propose a minimization criterion which takes account 
of the whole inertia tensor of each cluster.
\end{itemize}
The inertia tensor of a given cluster of points embedded in a $3D$ space 
is equivalent to the covariance matrix of their 
coordinates $(x,y,z)$. The covariance matrix $C$ reads :
\be
C = \left(
     \begin{array}{ccc}
      \sigma_x^2 & {\rm cov}(x,y)   & {\rm cov}(x,z)   \\
      {\rm cov}(x,y)   & \sigma_y^2 & {\rm cov}(y,z)   \\
      {\rm cov}(x,z)   & {\rm cov}(y,z)   & \sigma_z^2
     \end{array}
    \right)
\ee
where the diagonal terms are the variances of the variables $x, y$ and $z$, 
while the off-diagonal terms are the  covariances of the pairs of variables.
Diagonalizing  the covariance matrix provides a set of three eigenvalues 
and their associated eigenvectors. The largest 
eigenvalue $\lambda_1^2$ provides an information on the largest dimension 
of the cluster (thereafter considered as its length), and 
its associated eigenvector $\vec{u_1}$ provides the direction 
along which the length is striking. The second largest eigenvalue $\lambda_2^2$
(and its eigenvector $\vec{u_2}$) gives the same kind of information for
the width of the cluster, while the smallest eigenvalue $\lambda_3^2$
(and its eigenvector $\vec{u_3}$) gives an information on the 
thickness of the cluster. If we now consider
that data points cluster around a fault, eigenvalues and eigenvectors 
provide the dimensions and orientations of the fault plane,
the third eigenvector being normal to the fault plane.

The relationship between an eigenvalue and the dimension of the 
fault along the associated direction depends on the specific 
distribution of points on the plane. In the following, we shall assume 
for simplicity that data points are distributed uniformly
over a plane, so that the length of the fault $L$ is $\lambda_1\sqrt{12}$, and 
its width $W$ is $\lambda_2\sqrt{12}$. 
The square root of the third eigenvalue is the standard deviation 
of the location of events perpendicularly to the fault plane. If the fault
plane really represents the fault associated with the earthquakes, the
standard deviation  of the location of events perpendicularly to the fault plane
should be of the order of the localization uncertainty.
Our idea is to partition the data points by minimizing the sum of all 
$\lambda_3^2$ values obtained for each cluster, so that
the partition will converge to a set of clusters that tend to be 
as thin as possible in one direction, while being of arbitrary dimension
in other directions. This procedure defines a set of fault-like objects
in the most natural way possible. For each cluster, the knowledge of the 
third eigenvector $\vec{u_3}$
is then sufficient to determinate the strike and dip of the fault.

\section{A new method: the 3D Optimal Anisotropic Dynamic Clustering}

\subsection{Definition of the 3D Optimal Anisotropic Dynamic Clustering}

The general problem we have to solve is to partition a set of earthquake 
hypocenters into separate clusters labelled as faults. 
The previous section presented the general method to determine 
the size of a cluster as well as a new minimization criterion for 
anisotropic dynamic clustering. We shall now give a brief overview 
of a new algorithm that we propose, which also performs the 
automatic determination of the number of clusters to be used in the partition. 
This last property is particularly important in 
order for the fault network to be entirely deduced from the data with no a priori bias. 
To simplify at this stage of development (this can be refined in the future), 
we consider that the localization uncertainty of  earthquakes is uniform 
in the whole catalog and equal to $\Delta$. The idea is thus that all 
clusters should be characterized by $\lambda_3 < \Delta$. 
If not, part of the variance may be explained by something else than 
localization errors, {\it i.e.} another fault.
The algorithm can be described as follows.
\begin{enumerate}
\item We first consider $N_0$ faults with random positions, orientations and sizes.
\item For each earthquake in the catalog, we search for the closest fault and associate the former to the latter.  This provides a partition of seismic events in clusters.
\item For each cluster, we compute the position of its barycenter as well as its covariance matrix. This matrix is diagonalized,  which provides its eigenvectors 
and eigenvalues. From those last parameters, we compute the strike, the dip, 
the length and the  width, {\it i.e.} the characteristics of the fault which 
explain best the given cluster. The center of the fault is located at the 
barycenter of the cluster.
\item If we have $\lambda_3 < \Delta$ for all clusters, the computation stops, 
as the dispersion of events around faults can be fully explained by
localization errors. If not, we get back to step (2) and loop until we converge to a fixed geometry. Then, if  there is at least one cluster for which $\lambda_3 > \Delta$, we get to step (5).
\item We split the thickest cluster (the one which possesses the 
largest $\lambda_3$ value) by removing the associated fault and 
replacing it by at least $2$ other faults with random positions 
within that cluster. Obviously the cluster with the largest $\lambda_3$ 
is the one which needs  most to be fitted using more faults. Note that this step increases the total number of fault $N_0$ used in the partition.
\item We go back to step (2).
\end{enumerate}
The computation then stops when there are enough faults in the system 
to split the set of earthquake hypocenters in clusters that all obey the condition
$\lambda_3 < \Delta$. In addition, we remove from further analysis
those clusters which contain too few events or 
the earthquakes which are spatially isolated.

\subsection{Tests on synthetic datasets}

This algorithm has been tested on synthetic sets of events. The
synthetic catalog of earthquakes has been constructed as follows. We first choose 
the number of faults as well as their characteristics. 
We then put at random some data points on those planes, and add some 
noise to simulate the localization uncertainty characterizing the location
of natural events.  This set of data points is then used as an input for the new algorithm
presented in the previous section. Our goal is to check if 
the algorithm is able to find the characteristics of the 
original fault network. As we want to ensure that no a priori knowledge of the dataset can contaminate the determination of the 
clusters, we start the algorithm with an initial number of faults equal to $N_0=1$. 

We shall begin by detailing the example shown in Fig. \ref{synth_faults_trio}a. 
This dataset has been generated using two  parallel planes, each of them
being normal to a third one. All planes are vertical. Two planes strike $N90$ (and are located
$12km$ apart) while the third one  strikes $N0$. All planes are $20km$ 
long and $10km$ wide. Two hundred earthquakes are located randomly 
on each of those three planes while a random 
white noise in the interval $[-0.01km;+0.01km]$ is added to each coordinate of every earthquake. We set $\Delta=0.01km$ and start the 3D Optimal Anisotropic Clustering
algorithm on this dataset. As we start with $N_0=1$, the algorithm first fits 
the dataset with a single plane determined from the covariance matrix 
of the whole catalog, giving the solution shown in Fig. \ref{synth_faults_trio}b. 
Note that the best fitting plane is horizontal, and thus does not reflect at all the original geometry. Its length is $19.667$ km and its width is $16.798$ km.
For that plane, we find $\lambda_3 = 5.038$ km $> \Delta$, 
so that a second plane is introduced in order to decrease  $\lambda_3$.
The algorithm then converges to the fit proposed in Fig. \ref{synth_faults_trio}c. 
The plane $A$ strikes $N90$ and dips $S86$. Its length
is $16.985$ km and its width is $9.944$ km. Its $\lambda_3$ value is $2.933$ km. 
The plane $B$ strikes $N90$ and dips $N87$. Its
length and width are respectively $16.985$ km and $10.073$ km. Its $\lambda_3$ value is $3.043$ km. The two planes stand $11km$ apart.
The cluster associated to plane $A$ being the thickest (and its thickness being larger than $\Delta$), we split its plane in two, so that there are 
now three planes available to fit the data. After a few iterations, the algorithm 
converges the structure shown in Fig. \ref{synth_faults_trio}d. 
One can check that the spatial location of each plane is correct, while the azimuts and dips are determined within an error of less than $0.005$ degrees. The dimensions of the planes are determined within $3\%$. The value $\lambda_3$ for each plane is slightly lower
than $\Delta$. The algorithm has thus found by itself that it needed three planes to fit correctly the dataset, the remaining  variance being explained by the localization errors.

Using synthetic tests developed to demonstrate the power of the method, we 
considered several other fault geometries, with varying 
strikes, dips and dimensions. It is worth noting that the spatial extent of the 
reconstructed fault structure was found with great 
accuracy in each case. Our synthetic tests have been performed with a 
seismicity uniformly distribution on the fault planes. In the future, we plan to perform 
more in-depth tests with synthetic catalogs on more complex multi-scale 
fault networks in the presence of a possible non-uniform
complex spatial organization of seismicity, so as to better mimic real catalogs. 
But since, as we shall see in the next section, the extents of faults 
inverted form natural data sets seem reasonable as well (see next section), 
we conjecture that our method is robust with respect to the presence of
heterogeneity.

\subsection{Application to the Landers aftershocks sequence}

\subsubsection{Implementation}

As the method yields correct results on simple synthetic examples, 
we are encouraged to apply it to real time series, such as the aftershock sequence of 
the  1992 Landers earthquake in Southern California. We used the locations provided by 
{\it Shearer et al (2005)} and considered only the first 
two weeks of aftershock activity, {\it i.e.} a set of $3503$ events. 
We used such a limited subset of the full sequence due to limits set by
the required computation time. In the future, 
we plan to parallelize the algorithm in order to be able to deal with much
larger catalogs. The relative localization uncertainty is 
reported to be a few tens of meters y {\it Shearer et al (2005)}, but we set
$\Delta=1$ km to perform a first coarse scale inversion. Our justification
for setting $\Delta=1$ km is twofold. First, the absolute localization errors are
certainly larger than a few tens of meters. Second, using a value of 
$\Delta$ smaller than $1$ km  would imply to use a much larger catalog, with many more events to sample correctly small-scale features, hence increasing the computation time
prohibitively.  We detail hereafter the progression of the algorithm as the number of introduced faults increases, as we did in the previous 
section for the synthetic case shown in 
Fig. \ref{landers_fault_plate}a to \ref{landers_fault_plate}f. As the $3D$ spatial structure
of the sequence is quite complex, we show only the projection of the results on the horizontal plane.

Stage a) shows the result of the fit with a single fault, for which we
obtain $\lambda_3=2.9$ km. Since this value is larger than $\Delta=1$ km,
we introduce a second fault, and obtain
the pattern shown in stage b). The second fault is found to increase the quality 
of the fit mainly at the southern end of the
sequence, where a large amount of events are located. The largest $\lambda_3$ 
is still about $2.9$ km. We thus introduce
a third fault, with the optimal three fault structure shown in stage c). 
The new fault now helps to increase the quality of the fit in the northern end. 
Notice that this third fault is found to be horizontal, a signature of the competition
between several strands as observed in the synthetic test presented in
Fig. \ref{synth_faults_trio}b. Indeed, one can observe visually that this northern
region is characterized by at least three branches, which explains the 
optimization with an horizontal northern fault.
The value $\lambda_3$ for this last fault is about $2.5km$, leading to the introduction of
a fourth fault, whose optimal structure is shown in stage d). Now, the northern 
end is fitted more satisfactorily, and the largest $\lambda_3$ value drops 
to $1.8km$. The algorithm thus continues to introduce new faults, and we step directly to the pattern we obtain with $8$ faults which is shown in 
Fig. \ref{landers_fault_plate}e. At this stage, the algorithm has
dissected the central part of the dataset, but still does not
provide a good fit to the branch at $30km$ North of the origin. 
As the largest value $\lambda_3$ is $1.4km$, the algorithm still needs more faults.
Fig. \ref{landers_fault_plate}f shows the optimal fault structure fitting the 
data set of aftershock hypocenters with $12$ faults. The northern
part now appears to be fitted very nicely, while the southern part looks quite complex. 
As the largest value $\lambda_3$ is now $1.14km$, the algorithm needs more faults to fit the data with the threshold condition $\Delta=1$ km. 
We find that the largest value
$\lambda_3$ drops below $\Delta=1$ km for $16$ faults, for which the 
computation stops and yields the final pattern shown in 
Fig. \ref{landers_16fault}. When interpreting this figure, it is important to realize
that some fault planes are nearly vertical, which make them
barely visible in the projection shown in Fig. \ref{landers_16fault}.  

\begin{figure}
\begin{center}
\includegraphics[width=14cm]{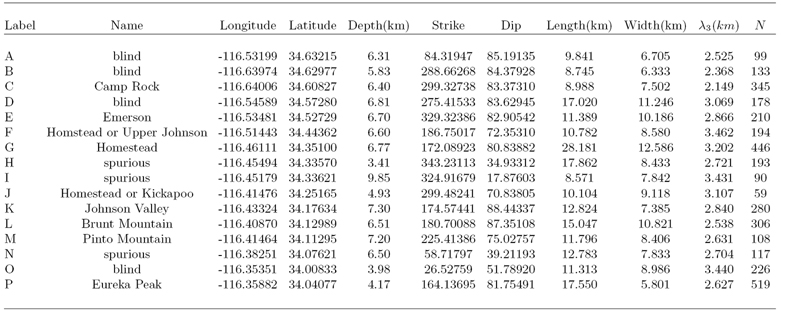}
\caption{\label{table_planes} Table of correspondances between fitting planes of the Landers, California, sequence and
faults in the same area. Each fitting plane is qualified by the label given on Fig. \ref{landers_16fault}. The name of
the corresponding fault is given in the second column. {\it Spurious} indicates that the plane certainly has no tectonic 
significance, while {\it blind} signifies that the fault doesn't intersect the free surface. Each plane is characterized
by the latitude, longitude and depth of its center, as well as its strike, deep and dimensions in $km$. The $2$ last columns
feature the $\lambda_3$ value of the corresponding cluster and the total number of events that belong to that cluster.
}
\end{center}
\end{figure}

\begin{figure}
\begin{center}
\includegraphics[width=14cm]{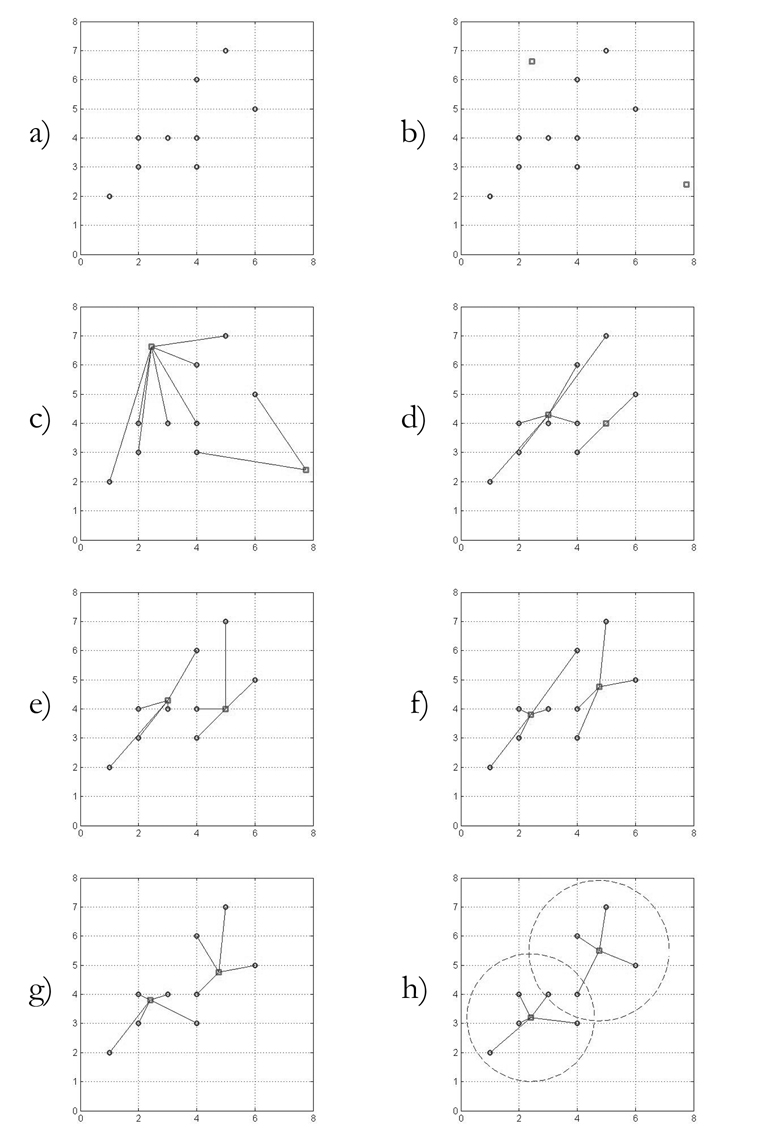}
\caption{\label{nudyn_plate_1} Presentation of the classical dynamic clustering method. 
a) set of $9$ data points represented by the circles which need 
to be partitioned into $2$ clusters - 
b) same as a) with the addition of the $2$ starting seeds
 for the clustering procedure (squares) - 
c) \& d) $1^{st}$ iteration (see main text) -
e) \& f) $2^{nd}$ iteration - 
g) \& h) $3^{rd}$ and last iteration - the
circles in dashed-line in h) give an idea of the size of clusters.
}
\end{center}
\end{figure}

\begin{figure}
\begin{center}
\includegraphics[width=14cm]{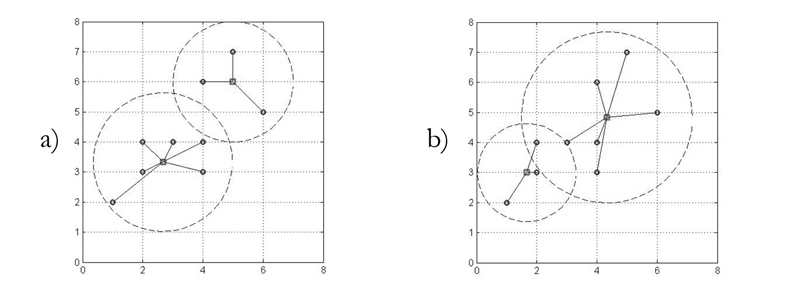}
\caption{\label{nudyn_plate_2} Final partitions for the same 
data set as in Fig. 1a) obtained for $2$ different starting 
configurations of the seed points (squares). 
The partition in a) corresponds to the global minimum of the sum 
of inertia over both clusters.
}
\end{center}
\end{figure}

\begin{figure}
\begin{center}
\includegraphics[width=14cm]{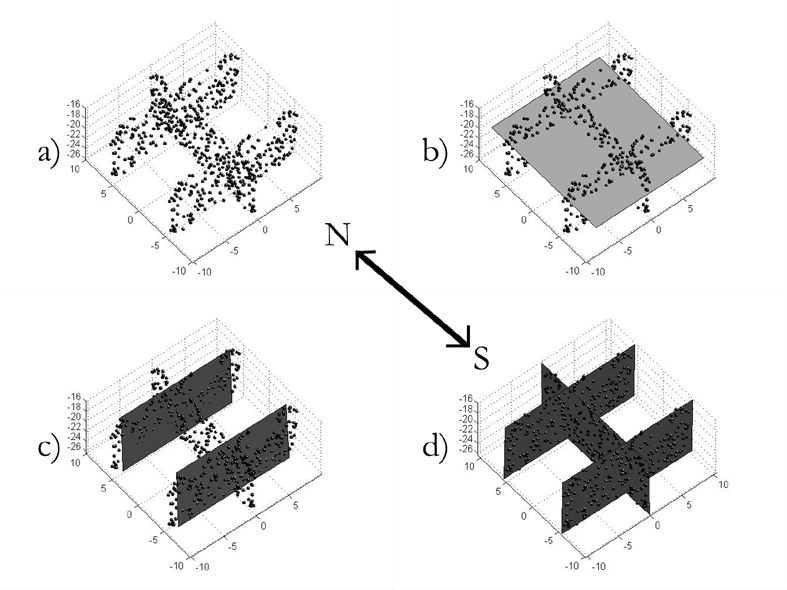}
\caption{\label{synth_faults_trio} Step by step progression of the optimal anisotropic dynamic clustering procedure proposed in this article, which is 
applied to a synthetic dataset.
Panel a) shows the $3D$ dataset -
Panel b) shows a fit with $1$ plane -
Panel c) shows the best fit with $2$ planes -
Panel d) shows the best fit with $3$ planes. In this last case, we recover the correct set of planes with which the data set
had been generated.
}
\end{center}
\end{figure}

In order to describe the fault network, it is convenient to label the $16$ faults
from $A$ to $P$, which allows us to discuss this pattern fault by fault. The parameters
of the $16$ fault planes (size and orientation) are given in Table \ref{table_planes}. 
These fault planes will now be classified into three
different catagories, namely (i) spurious planes (which have no tectonic significance) (ii) previously known planes (that correspond to mapped faults) (iii) 
unknown planes (that may correspond to blind faults
or to otherwise structures  unmapped for whatever reason).

\subsubsection{Spurious planes}

An inspection of Table \ref{table_planes} reveals that most
proposed planes dip close to the vertical. Three planes,
$H$, $I$ and $N$ have rather abnormal dips, which leads us to suspect 
that they are spurious. Indeed, $H$ and $I$ are near
normal to the plane $G$, which is located in a zone with rather diffuse 
seismicity in the direction normal to that plane. 
Introducing planes $H$ and $I$ is a
convenient way to reduce the variance in that zone. It is likely that those
planes have no tectonic significance, but have been found by the 
algorithm as the way to satisfy the criterion on $\lambda_3$
(we shall come back below to this argument). Plane $N$ also seems to be
introduced just to decrease the variance in a zone displaying fault branching.
All other $13$ fault planes out of the $16$ have dips larger than 
$50$ degrees, so that we have {\it a priori} no reason 
to remove them.

\subsubsection{Previously known faults}

Because the fault planes have been obtained by fitting seismicity data, none of them 
cross-cut the free surface. It is however interesting
to compare the planes in Fig. \ref{landers_16fault} to the faults mapped at the surface in the Landers area (see {\it Liu et al, 2003}), and to underline possible correspondances. 
For example, plane $C$ clearly corresponds to
the southern end of the Camp Rock fault. Plane $E$ corresponds to the Emerson fault. 
Plane $G$ is a good candidate for the Homestead Valley fault. 
Note that surface faulting is quite complex and diffuse in that zone. Continuing to the South,
planes $K$ and $L$ seem to match respectively with the Johnson Valley and Brunt Mountain faults, while plane $P$  is the Eureka Peak fault. All those faults are known to have been activated during the Landers event. Plane $F$ is located in quite a complex
faulting zone, but its azimuth fits well with both the Northern end of the 
Homestead Valley fault or with the Upper Johnson Valley fault.
In the former case, plane $G$ would then fit with most of the Homestead Valley fault and the Maume fault. Plane $J$ is also located in a zone of complex faulting, and it is difficult 
to guess if it corresponds to the Southern end of the Homestead 
Valley fault, or to the Kickapoo fault. Note that it may also feature events
from both faults.

\subsubsection{Unknown faults}

This last set contains planes $A$, $B$, $D$, $M$ and $O$. Planes $A$, $B$ and $D$ obviously fit blind faults, that one can guess, for instance, in Fig. \ref{longterm_landers} which plots the full set of events
in {\it Shearer et al (2005)}'s catalog in the same area. 
In the Southern end, plane $M$ may fit
with the Pinto Mountain fault, while $O$ would be a genuine blind fault.

\section{Discussion and Conclusion}

We have introduced a new powerful method, the 3D Optimal
Anisotropic Dynamic Clustering, that allows us to partition a $3D$
 set of earthquakes hypocenters into distinct
anisotropic clusters which can be interpreted as fault planes.
The method was first applied to a variety of synthetic data sets, which
confirmed its ability to recover correctly with high accuracy the 
existence of present faults. We then applied the 3D Optimal
Anisotropic Dynamic Clustering to the sequence of
aftershocks following the 1992 Landers event in California. In this later case, 
we were able to recognize the faults that were already known by surface 
mapping. In addition, the 3D Optimal
Anisotropic Dynamic Clustering method allowed us to identify
some additional blind faults. These faults appear to make sense when
looking at the long-term epicenter maps or $3D$ plots of seismicity. The advantage
of our method is that it finds automatically the characteristics of the 
fault planes as well as the hypocenters that belong to each
of them, without any need for the operator to pick them manually. 
The method provides the orientation of the faults,
as well as the size of their active part during the earthquake sequence
used to identify them.

The main drawback of this method is that it can propose fault planes that
are likely spurious (error of type II), as they lack any tectonic meaningful interpretation. 
This problem appears in the presence of diffuse seismicity, for which the
3D Optimal Anisotropic Dynamic Clustering method forces the introduction
of fault planes in the diffuse zone in order to reduce the global variance 
of the distances from earthquakes to the proposed fault planes. In the case
of the Landers aftershock sequence, we
were able to sort them out from their anomalous dip parameters, compared with
all known fault planes in the Landers area which are  nearly vertical, as also
confirmed by focal mechanisms. 
This post-processing step will become much more
subjective when trying to sort out spurious planes in more complex tectonic settings. 
We thus believe that
our procedure should use additional information available in earthquakes catalogs
than just the 3D hypocenter locations. For example, focal mechanism characteristics
are now determined for a large number of events, so that the inertia of a cluster 
should use not only the position of a given
event in the cluster, but also a criterion of compatibility of the focal mechanism of this event with the orientation of the best fitting plane of that same cluster. 
The idea is to make the algorithm
converge to a set of anisotropic clusters over which the distribution of focal mechanisms is approximately uniform. This extension of the method will be developed 
and tested elsewhere.

In the same vein, we propose another future extension of the
3D Optimal Anisotropic Dynamic Clustering consisting of using  
the information on waveforms contained in the Southern
California earthquakes catalog of {\it Shearer et al
(2005)}. The method used to build this catalog consists in grouping events according
to the similarity of their waveforms. This first step thus yields a proto-clustering 
of events. Time-delays of wave arrival
times for events belonging to the same proto-cluster yield relative locations 
of those grouped events. Events belonging to
the same proto-cluster display similar waveforms and are thought to be
characterized by similar rupture mechanisms, thus
may betray successive ruptures on the same fault. {\it Shearer et al (2005)} 
used this idea and performed principal component
analyses on each proto-cluster of the Northridge, California, 
aftershocks sequence. This idea is similar to ours,  but nothing proves that a single
proto-cluster samples only one fault segment. For example, when looking at the seismicity in the Imperial fault area, {\it Shearer (2002)} has been able to identify
some very fine-scale features in earthquakes clouds that otherwise looked 
very fuzzy. In this last example, a proto-cluster of events was
found to be clearly composed of two parallel lineaments, with a $500m$ offset between them. However, the operator still has to use a ruler and a protractor to compute the
dimension and orientation of each lineament, and his eyes to count them. 
Using a method such as ours on that specific proto-cluster would naturally
partition it into two (sub)clusters, and provide their associated 
dimensions and orientations. Thus, a natural future application of our method
would be to apply it on each of the clusters resulting from {\it Shearer et al
(2005)}'s proto-partition. The advantage is that the number of events
in each cluster is small, so that the algorithm will converge very quickly, and that events will be already pre-sorted according to the similarity of their rupture mechanism. The fact that all events within proto-clusters are located
with an accuracy of the order of few tens of meters should allow one to provide 
a very detailed fit of the $3D$ spatial  structure of the catalog. 

\begin{figure}
\begin{center}
\includegraphics[width=14cm]{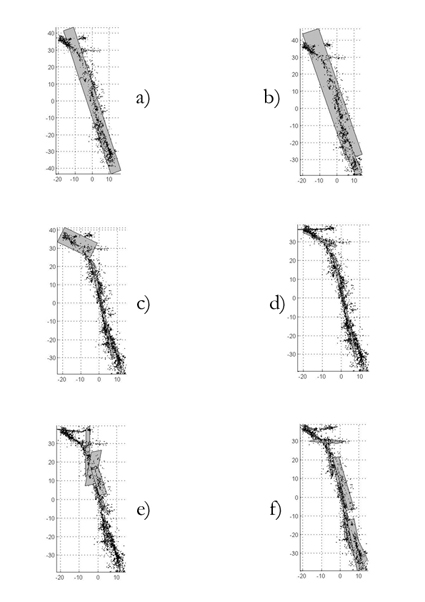}
\caption{\label{landers_fault_plate} Progression of our method of dynamic 
clustering when applied to a subset
of the 1992 Landers, California, aftershocks sequence.
a) fit of the sequence with $1$ plane -
b) fit with $2$ fault planes -
c) fit with $3$ fault planes -
d) fit with $4$ fault planes -
e) fit with $8$ fault planes -
f) fit with $12$ fault planes.
}
\end{center}
\end{figure}

\begin{figure}
\begin{center}
\includegraphics[width=14cm]{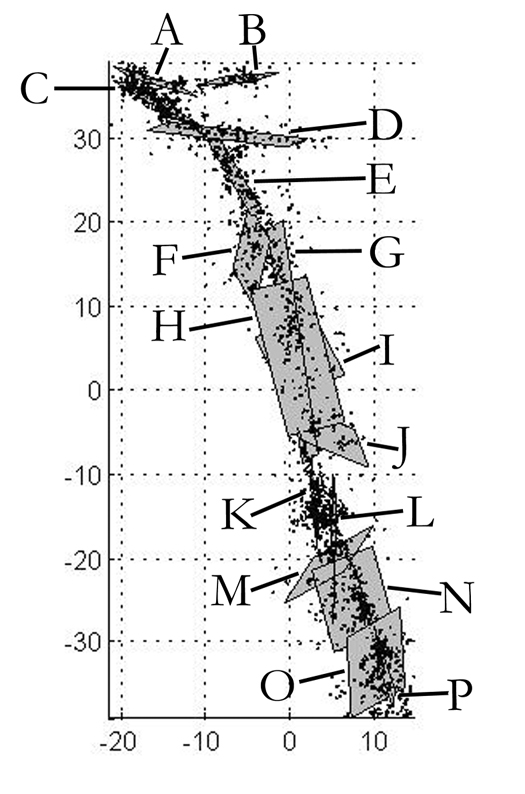}
\caption{\label{landers_16fault} Fit with $16$ fault planes of the 1992 Landers, California, 
aftershock data set. The value $\lambda_3$
of each cluster is smaller than the imposed resolution threshold $\Delta=1$ km.
}
\end{center}
\end{figure}

\begin{figure}
\begin{center}
\includegraphics[width=14cm]{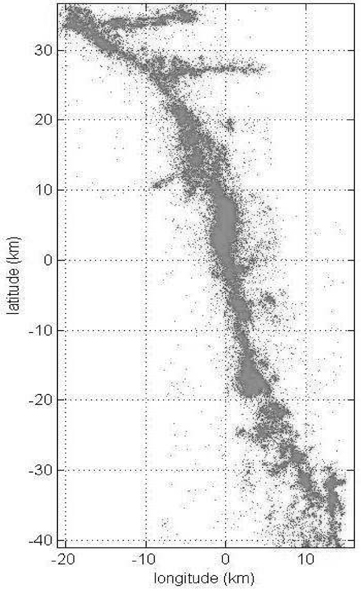}
\caption{\label{longterm_landers} Epicentral view of the full catalog available for the Landers
earthquake.
}
\end{center}
\end{figure}

We would also like to stress that the 3D Optimal
Anisotropic Dynamic Clustering method provides a natural tool for 
understanding the multiscale structure
of fault zones. In a recent paper, {\it Liu et al (2003)} showed that the 
spatial dispersion of aftershocks epicenters
normal to the direction of Landers main rupture could not be explained 
by localization errors alone, betraying the complex 
mechanical behaviour of this fault zone. In the examples
presented in this paper, we considered that the resolution parameter $\Delta$ 
controlling the number of faults necessary to explain the seismic data 
could be mapped in a natural way on the spatial localization uncertainty of the spatial 
location of the events. While this is a natural choice, more generally,
$\Delta$ can take any arbitrary value assigned by the user and, in such a case,
it must be interpreted as the spatial resolution at which the user
wants to approximate the anisotropic fault structure defined by the catalog of events.
For example, if we consider the choice $\Delta=10km$, 
the output is the same as the one presented on Fig. \ref{landers_fault_plate}a, as
only one fault is necessary to fit the data at this coarse resolution. 
As we make $\Delta$ decrease, the number of necessary 
faults increases as we have seen. Fig.~7 shows  that the 
number of fault planes necessary to fit the Landers' data varies with
$\Delta$ as 
\be
N(\Delta) \simeq \Delta^{-\mu}~, \quad {\rm  with} \quad \mu=2~,
\ee
over a limited range. For very large $\Delta$'s, it is obvious
that just one fault is necessary. This
powerlaw behaviour is of course reminiscent of a scale-invariant
geometry, but we still have to clarify a few points before any further serious conclusion. 
The first one is a theoretical one: usual methods used to compute the
fractal dimension of a given object consider scale-dependent approximations 
of that object with isotropic balls (for example,
box-counting uses isotropic balls with the $L_1$ norm, while the correlation method uses isotropic balls with the $L_2$ norm). In the present
case, we approximate our dataset by highly anisotropic ``balls,'' which
in addition do not have the same size. It thus seems inappropriate to
interpret $\mu$ as a genuine fractal dimension. The second is a pragmatic one: 
for a given scale-invariant distribution of points,
the measured fractal dimension depends on its cardinal number. 
The reason is that, the largest is the cardinal,
the better the distribution is sampled, and the better is its statistical characterization. If the cardinal is low, the multi-scale
structure of the distribution is not properly sampled and the measured fractal dimension is biased. As we still do not know
the effect of the bias that may affect the determination of the
exponent $\mu$, much more work is needed in that direction.

Considering earthquake forecasting and seismic hazard assessment, the 
knowledge of the fault network architecture in the vicinity of large events will
also help to test different competing hypothesis on 
stress transfer mechanism. Until now, two strategies have been used 
in order to check for the 
consistency between the geometry of the main shock rupture and 
the spatial location of aftershocks. The first one considers that
all rupture planes of aftershocks are parallel to the main rupture. The second
one postulates that the rupture planes are optimally oriented.
None of those hypotheses turn out to be true (see {\it Steacy et al (2005)}).
Therefore, we think that imaging accurately the $3D$ structure 
of a fault network constitutes 
one of the most important steps in order to decipher static or dynamic earthquake 
triggering. The 3D Optimal
Anisotropic Dynamic Clustering provides a first significant step in this direction.



{}

\end{article}

\newpage

\newpage

\newpage

\newpage

\newpage

\newpage

\newpage

\begin{figure}
\begin{center}
\includegraphics[width=14cm]{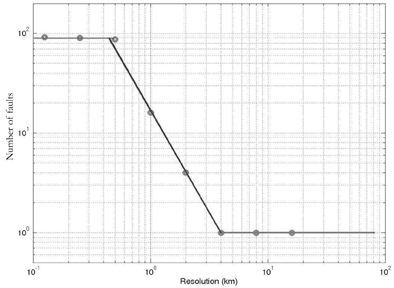}
\caption{\label{fractal} Number of fault planes necessary to fit the set of aftershocks
of the Landers earthquake, as a function of the resolution $\Delta$.
Note the finite range of scales over which the power law
behavior  $N(\Delta) \simeq \Delta^{-2}$ is valid.
}
\end{center}
\end{figure}

\end{document}